\documentclass[reprint,
 superscriptaddress,
 amsmath,
 amssymb,
 aps,
]{revtex4-1}
\usepackage[utf8]{inputenc}
\usepackage{graphicx}
\usepackage{dcolumn}
\usepackage{bm}
\usepackage{natbib}
\usepackage{dirtytalk}

\begin{document}

\title{Non-destructive X-ray imaging of patterned delta-layer devices in silicon}

\author{Nicolò D'Anna}
\email{nicolo.danna@psi.ch}
\affiliation{Paul Scherrer Institut, 5232 Villigen, Switzerland}
\affiliation{Department of Physics and Quantum Center, Eidgenössische Technische Hochschule Zürich, CH-8093 Zürich, Switzerland}

\author{Dario Ferreira Sanchez} 
\affiliation{Paul Scherrer Institut, 5232 Villigen, Switzerland}

\author{Guy Matmon}
\affiliation{Paul Scherrer Institut, 5232 Villigen, Switzerland}

\author{Jamie Bragg}
\affiliation{London Centre for Nanotechnology, University College London, WC1H 0AH, London, UK}
\affiliation{Department of Electronic and Electrical Engineering, University College London, London WC1H 0AH, UK}

\author{Procopios~C.~Constantinou} 
\affiliation{Paul Scherrer Institut, 5232 Villigen, Switzerland}
\affiliation{London Centre for Nanotechnology, University College London, WC1H 0AH, London, UK}
\affiliation{Department of Physics and Astronomy, University College London, WC1E 6BT, London, UK}

\author{Taylor J.Z. Stock}
\affiliation{London Centre for Nanotechnology, University College London, WC1H 0AH, London, UK}

\author{Sarah Fearn}
\affiliation{London Centre for Nanotechnology, University College London, WC1H 0AH, London, UK}
\affiliation{Department of Materials, Imperial College of London, London SW7 2AZ, UK}

\author{Steven R. Schofield}
\affiliation{London Centre for Nanotechnology, University College London, WC1H 0AH, London, UK}
\affiliation{Department of Physics and Astronomy, University College London, WC1E 6BT, London, UK}

\author{Neil~J.~Curson}
\affiliation{London Centre for Nanotechnology, University College London, WC1H 0AH, London, UK}
\affiliation{Department of Electronic and Electrical Engineering, University College London, London WC1H 0AH, UK}

\author{Marek~Bartkowiak}
\affiliation{Paul Scherrer Institut, 5232 Villigen, Switzerland}

\author{Y. Soh}
\affiliation{Paul Scherrer Institut, 5232 Villigen, Switzerland}

\author{Daniel Grolimund}
\affiliation{Paul Scherrer Institut, 5232 Villigen, Switzerland}

\author{Simon Gerber}
\affiliation{Paul Scherrer Institut, 5232 Villigen, Switzerland}

\author{Gabriel Aeppli}
\email{gabriel.aeppli@psi.ch}
\affiliation{Paul Scherrer Institut, 5232 Villigen, Switzerland}
\affiliation{Department of Physics and Quantum Center, Eidgenössische Technische Hochschule Zürich, CH-8093 Zürich, Switzerland}
\affiliation{Institute of Physics, Ecole Polytechnique Fédérale de Lausanne (EPFL), 1015 Lausanne, Switzerland}    

\date{\today}

\begin{abstract}
\textbf{The progress of miniaturisation in integrated electronics has led to atomic and nanometre-sized dopant devices in silicon. Such structures can be fabricated routinely  by hydrogen resist lithography, using various dopants such as phosphorous and arsenic. However, the ability to non-destructively obtain atomic-species-specific images of the final structure, which would be an indispensable tool for building more complex nano-scale devices, such as quantum co-processors, remains an unresolved challenge. 
Here we exploit X-ray fluorescence to create an element-specific image of As dopants in silicon, with dopant densities in absolute units and a resolution limited by the beam focal size (here $\sim1~\mu$m), without affecting the device's low temperature electronic properties. 
   The As densities provided by the X-ray data are compared to those derived from Hall effect measurements as well as the standard non-repeatable, scanning tunnelling microscopy and secondary ion mass spectroscopy, techniques. Before and after the X-ray experiments, we also measured the magneto-conductance, dominated by weak localisation, a quantum interference effect extremely sensitive to sample dimensions and disorder. Notwithstanding the $1.5\times10^{10}$~Sv ($1.5\times10^{16}$ Rad/cm$^{-2}$) exposure of the device to X-rays, all transport data were unchanged to within experimental errors, corresponding to upper bounds of 0.2 Angstroms for the radiation-induced motion of the typical As atom and 3$\%$ for the loss of activated, carrier-contributing dopants.  With next generation synchrotron radiation sources and more advanced optics, we foresee that it will be possible to obtain X-ray images of single dopant atoms within resolved radii of 5~nm.}
\end{abstract}
\maketitle

The ability to build nanometre-scale dopant structures buried in silicon has led to great progress in classical and quantum technologies \cite{silicon_quantum_electronics}.  As the patterned structures become increasingly small and complex, it becomes indispensable to develop techniques to image the dopant structures non-destructively for device inspection and quality control \cite{device_metrology,3D_imaging,high_res_3d_imaging}. 

Scanning tunnelling microscopy (STM) can be used to pattern acceptors and donors into silicon with atomic resolution using hydrogen resist lithography \cite{Dwyer:2021aa,PhysRevLett.91.136104}. The technique has been used to create complementary metal-oxide-semiconductor compatible structures, including two-dimensional conductive sheets \cite{siP_sheet}, three-dimensional structures \cite{3D_strucutures}, nano-wires \cite{1D_wire}, and quantum dots \cite{1D_dot}. 
Precisely measuring the location of buried dopants patterned by STM is challenging and can only be accomplished with STM itself for patterns extremely near to the surface \cite{PhysRevB.95.075408,Voisin:2021aa}.
Techniques capable of imaging such nano-scale structures are typically destructive, such as secondary-ion mass spectrometry (SIMS) \cite{SIMS} and atom probe tomography \cite{atome_probe}, making them unsuitable for device quality control \cite{metrology_and_failure}.
Two techniques that can image the dopants non-destructively are broadband electrostatic force microscopy (bb-EFM) \cite{bb_EFM} and infrared ellipsometry \cite{infrared_elipsometry}, however both come with limitations. In particular, \mbox{bb-EFM} can only measure the polarity of the dopant and not its elemental species, whereas infrared ellipsometry can, in principle, obtain information regarding the species and density of atoms, but it is model-dependent and requires elaborate fits to the data.

Here we show that X-ray fluorescence (XRF) can be used to create non-destructive atomic-species specific images of dopants in silicon with a resolution only limited by the beam-size, in our case of order one micron.
This technique uses synchrotron X-rays to locally ionise the atoms in the investigated device, leading to the emission of photons via fluorescence. The measurements are conducted at ambient temperature and pressure, and the photon spectrum is analysed to obtain the species and densities of the atoms in the device. Low-temperature magneto-transport on the two-dimensional Hall-bar device before and after imaging with the X-ray fluorescence demonstrates that the technique does not alter the electrical characteristics of the device, namely the free carrier density, electron mean free path, coherence length, and vertical confinement. We conclude therefore that the technique is non-destructive. As an extension of the principle demonstrated here, by rotating the sample in the X-ray beam it will be straightforward to obtain a tomographical three-dimensional reconstruction of the atoms' positions in the device \cite{high_res_3d_imaging,3D_tomo}.

\section*{X-ray fluorescence} 
When an X-ray photon impinges on an atom it can be absorbed by the atom that will, in turn, be ionised. Inner orbital electrons are expelled from the atom and replaced by outer orbital electrons. In this process photons are emitted with wavelengths corresponding exactly to the energy difference between the electrons' orbitals. Therefore, the resulting energy spectrum of the fluorescence photons will uniquely identify the atomic species of the ionised atom. In the presence of many different atoms the fluorescence spectrum will be the sum of the different spectral lines. As each atomic spectrum is well known, it is straightforward to decompose an arbitrary fluorescence spectrum into element-specific components \cite{Adams2011}.

\begin{figure*}[t]
  \centering
      \includegraphics[width=\linewidth,trim=0cm 2cm 0cm 2cm, clip=true]{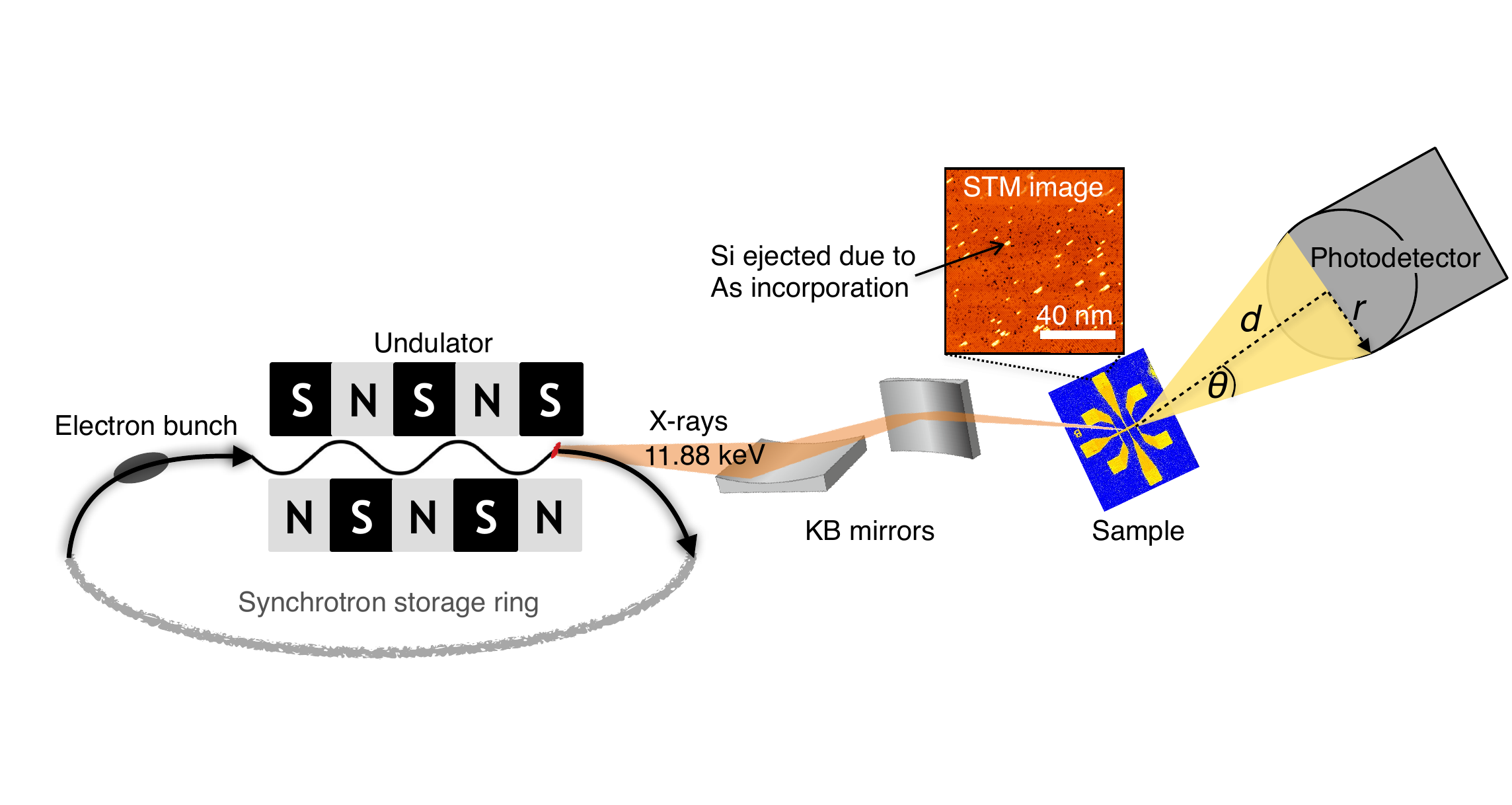}
     \caption{\textbf{Schematic of the X-ray fluorescence measurement}. From left to right: Electron bunches from the synchrotron are directed through the undulator magnets emitting highly collimated photons on account of the repeated electron beam bending.
    The X-ray beam with a photon energy of 11.88~keV from the synchrotron is focused to $1~\mu$m with the help of a Kirkpatrick-Baez (KB) mirror system. Fluorescence photons from the sample that is illuminated by the X-ray beam are emitted in all directions and a photodetector is placed at $d=2$~cm away from the sample, collecting the photons from a solid-angle $\Omega=2\pi(1-\cos\theta)=2\pi(1-d/\sqrt{d^2+r^2})=0.04\pi$.
    The inset shows an STM image of the sample's doped silicon surface before silicon overgrowth; the short bright lines are rows of Si dimers ejected from the surface plane due to the incorporation of As atoms.}
    \label{fig_setup}
\end{figure*} 

XRF experiments reported here were conducted at the microXAS beamline of the Swiss Light Source (SLS) synchrotron \cite{microXAS}.
The beamline produces high brightness X-rays in the energy range from $\sim4$ to 22~keV. At the photon energy of $11.88$~keV used here, the delivered photon flux is approximately $I_0$=10$^{10}$~photons/sec when the beam is focused to $1\times1$~$\mu$m using a Kirkpatrick–Baez mirror system, and an energy resolution of $\Delta E/E <10^{-4}$ is chosen.
The beam was set to normal incidence. An X-ray energy of 11.88~keV is sufficiently high to dislodge core electrons from the As $K$-edge, without exciting the gold atoms found in parts of the sample holder.

Measurements are conducted in air at room temperature, with a gentle flow of helium gas into a 15~mm long pinhole cavity that encapsulates the silicon drift detector (SDD), with the exit gas flow located 2~mm from the sample position. A silicon drift detector with an active area of 50~mm$^2$ is placed in close proximity of the sample to maximise fluorescence photon collection.  Figure~\ref{fig_setup} shows a schematic of the experiment; the solid angle captured by the detector is $\Omega=0.04\pi$.
The detector not only measures the intensity of the fluorescence photons but also resolves their energy spectrum, \textit{i.e.}, it counts the number of photons reaching the detector as a function of photon energy, as shown in Fig.~\ref{fig_fluorescence}e. The collected spectrum is then decomposed into the sum of the individual atom-specific spectra with the PyMca software \cite{PyMca}.
To determine the atom density from the detected fluorescence, it is compared to an arsenic containing reference sample from \b{nanoXRF standards} \cite{nanoXRF_standard} with a known density under the same X-ray beam illumination and placed at the same position as the measured device.
The intensity of each fluorescence peak in the spectrum depends not only on the density of atoms participating in the fluorescence process, but also on the ionisation cross-section. Note that these X-ray ionisation cross-sections are well-known and do not depend on factors such as the atom's depth or environment.
In XRF the measured density corresponds to the absolute number of atoms, unlike other non-destructive imaging techniques which measure only electrically activated dopants \cite{device_metrology}. By comparing the atomic density to the free carrier density (obtained from magneto-transport, see Magneto-Conductance section) it is thus possible to deduce the activation percentage in a given device. Knowing the dopant electrical activation is important for optimising device fabrication; in particular when making atomic-scale devices it is important to have an activation percentage close to 100$\%$ to ensure that all donors contribute an electron to the conduction band.

At the microXAS beamline, the beam position fixed and the sample was swept across the beam with a step size of 500~nm, and for each position a full spectrum of the fluorescence was recorded. The data collected in this way contain the information of the atomic concentrations at each position of the scan, from all elements that are excited with the chosen X-ray energy. By decomposing the full spectrum at each position into a sum of spectra from each possible element, a two-dimensional elemental density map is obtained. 

\section*{Subsurface imaging} %
\begin{table*}
\centering
\begin{tabular}{c|cccccc}
Device~& ~$n_{\rm STM}$ (10$^{14}$ cm$^{-2}$)~ & ~$n_{\rm XRF}$ (10$^{14}$ cm$^{-2}$)~ & ~$n_{\rm Hall}$ (10$^{14}$ cm$^{-2}$)~ & ~$n_{\rm SIMS}$ (10$^{14}$ cm$^{-2}$)~ & ~$t_{\rm SIMS}$ (nm)~ & $t_{\rm MR}$ (nm)  \\
 \hline
\#1 & $1.6\pm0.3$  & $1.40\pm0.07$ & $1.31\pm0.03$ &  $1.8\pm0.2$ & $2.7\pm0.2$  & $0.97\pm0.02$\\
\#2 & $0.10\pm0.03$  & $0.06\pm0.01$ & \textendash\textendash\textendash &  $0.21\pm0.02$ & $3.6\pm0.4$  & \textendash\textendash\textendash \\
\end{tabular}
\caption{\textbf{Dopant density and layer thicknesses.} As density $n$ of the two devices measured with STM, XRF, Hall effect and SIMS. Additionally the As layer thickness $t$ is given as measured by SIMS and MR. Device \#2 was not conductive and, therefore, Hall measurements were not possible.}
\label{table_1}
\end{table*}
Structures consisting of atomically thin layers of As (‘As $\delta$-layers’) buried 30~nm below the Si(100) surface were patterned into $20\times200~\mu$m$^2$ Hall-bars and contacted with aluminium as detailed in the methods.
The As layer is made by exposing atomically flat silicon to a dose of arsine, annealing the wafer to incorporate the As into the surface layers, then overgrowing with epitaxial silicon. The As density is simply controlled by the total As dose. Two such structures are studied here, one with a nominal As density of \mbox{$n_{\rm As}=1.6\times10^{14}$~cm$^{-2}$} and the other with \mbox{$n_{\rm As}=1\times10^{13}$~cm$^{-2}$}. The dopant density is determined with the STM before the silicon overgrowth by counting Si atoms ejected by the incorporated As, as seen on the inset in Fig.~\ref{fig_setup} and explained in \cite{Stock_As_count} and the methods.
The same devices were used for the XRF and the magneto-resistance (MR) measurements.
Table~\ref{table_1} summarises the devices' density and thickness as measured by STM, XRF, SIMS and MR.
Additionally, to quantify the background As dopant density a reference sample was measured.

\begin{figure*}[t]
  \centering
      \includegraphics[width=\linewidth,trim=2.5cm 0cm 1.5cm 0cm, clip=true]{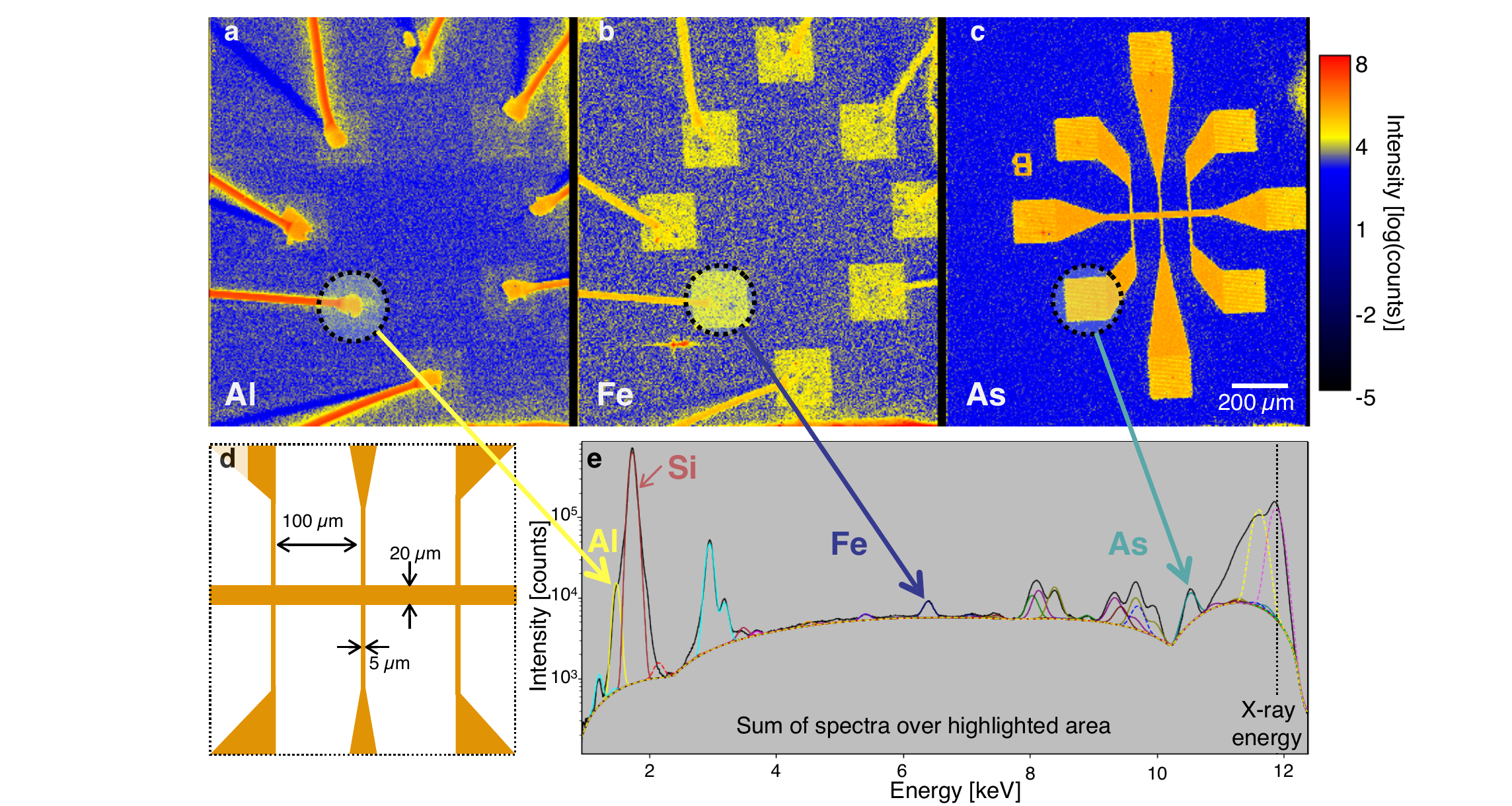}
     \caption{\textbf{X-ray fluorescence image of an As Hall-bar device.} \textbf{a}-\textbf{c} Al, Fe, and As distribution of device \#1 with a density of $n_{\rm As}=1.4\times10^{14}$~cm$^{-2}$ for a photon energy of 11.88~keV, a beam size of $1\times1$~$\mu$m$^2$ and a step-size of 0.5~$\mu$m. At each step a spectrum is recorded during 200~ms. \textbf{d} Sketch of the top view of the samples with the As Hall-bar structure shown in orange. \textbf{e} Sum of the fluorescence spectra taken at each scan point within the highlighted area in \textbf{a}-\textbf{c}. The black line represent the measured data and the coloured lines are fits to individual elements. See Fig.~\ref{fig_spectrum} in the methods section for details on the elemental contributions.}
    \label{fig_fluorescence}
\end{figure*} 

Figure \ref{fig_fluorescence}d shows a sketch of the studied devices, where the orange colour illustrates the two-dimensional As layer.
Figures \ref{fig_fluorescence}a-c depict the higher density Hall-bar structure \#1 as imaged by XRF for Al, Fe, and As, respectively. For each pixel of the image a spectrum is recorded for a duration of 200~ms. The sum of many such spectra is shown in Fig.~\ref{fig_fluorescence}e with fitted peaks to deduce the elemental origin.
Note that each element's fluorescence spectrum has peaks at unique energies (see methods), such that fitting the data is straightforward.

The As image in Fig.~\ref{fig_fluorescence}c clearly shows the conductive layer of interest, which defines the Hall-bar and its contact leads. 
The unique possibility to distinguish different atomic species makes it possible to verify whether there is contamination in the device. Here the spectrum contains traces of many elements (see methods section for element identification), which originate from the lead-less chip-carrier, the glue used to fix the sample and the He gas that is blown on the sample (traces of Ar in Fig.~\ref{fig_spectrum}). We also see that the Al contact pads and bonding wires contain not only aluminium, but also a very small quantity of Fe. The exact density is obtained by comparing the intensity of the fluorescence to the reference sample. While the Fe density is only $n_{\rm Fe}=1\times10^{13}$~cm$^{-2}$ in the Al contacts, it provides a stronger XRF signal than the aluminium whose density is $n_{\rm Al}=6.0\times10^{15}$~cm$^{-2}$. This is due to the larger (107.3~cm$^{-2}$/g) absorption cross-section of iron, compared to aluminium (14.76~cm$^{-2}$/g), and the considerably larger absorption of the low energy Al fluorescence by the air/He atmosphere and by the detector window . 
The As density is uniform across the entire Hall-bar and is found to be $n_{\rm As}=1.4\times10^{14}$~cm$^{-2}$ and $n_{\rm As}=5.6\times10^{12}$~cm$^{-2}$, for the two devices measured. 
For both the values obtained with XRF and STM agree within the uncertainty (see Tab.~\ref{table_1}). The uncertainty in XRF measurements is low because atomic cross-sections are well known such that the reference sample yields an uncertainty of less than $5\%$.

By comparing the atomic density to the free carrier density obtained from Hall measurements we find that the dopant activation in the high-density Hall-bar \#1 is \mbox{$94\pm5\%$}. The low-density device \#2 was not conductive and no Hall density could be measured.
\begin{figure*}[t]
  \centering
      \includegraphics[width=\linewidth, trim=0cm 2cm 0cm 2.5cm, clip=true]{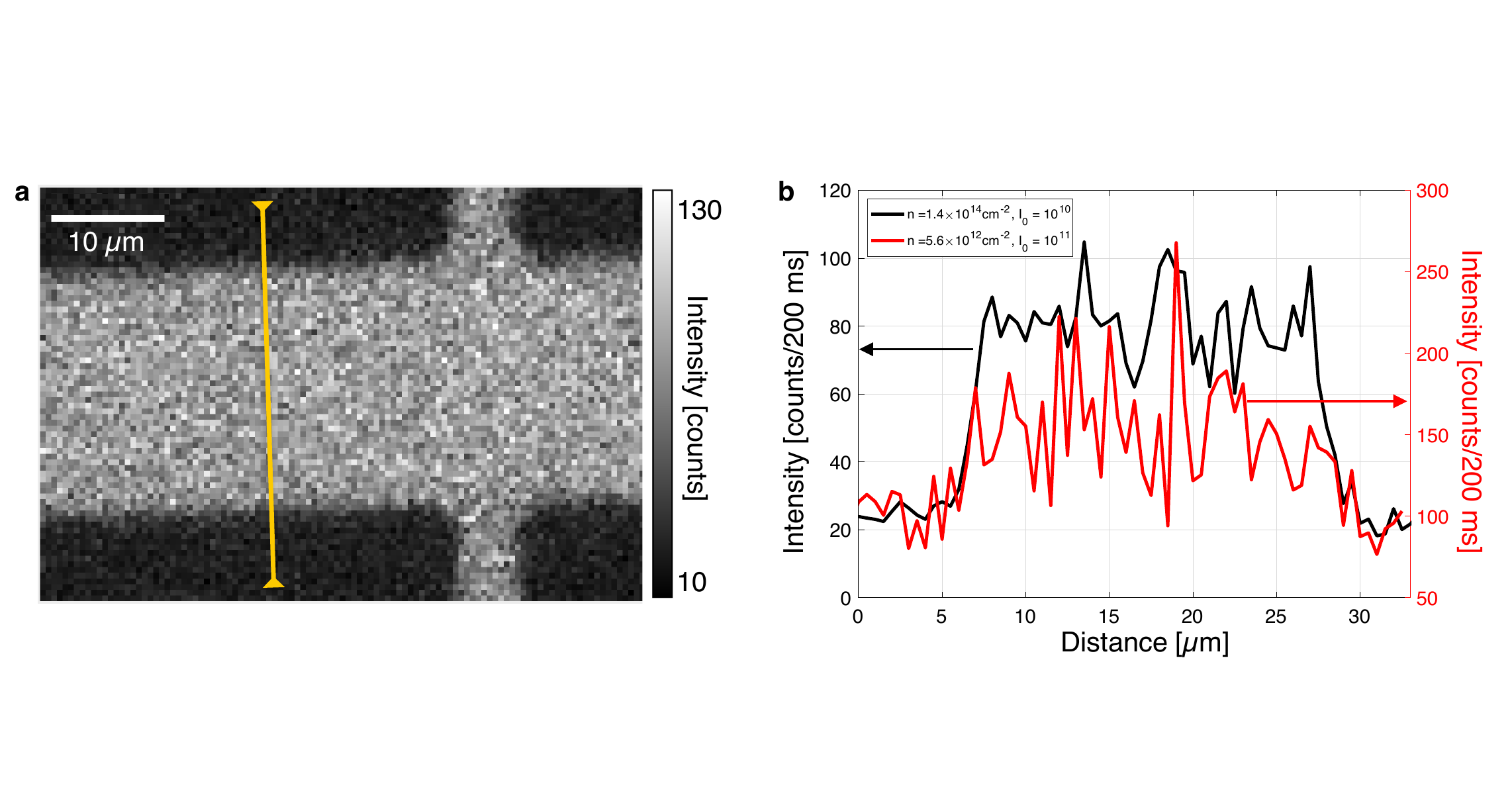}
     \caption{\textbf{Fluorescence contrast across the Hall-bars. a} As fluorescence image of the high density Hall-bar device \#1, taken with a beamsize of $1\times1$~$\mu$m$^2$ and an averaging time of 200~ms. The image is obtained by scanning the beam across the sample, with the x-axis parallel and the y-axis perpendicular to the Hall-bar, respectively. The yellow line denotes the line trace. \textbf{b} Line traces in counts per dwell time measured across the Hall-bars of the high-density (black, 1$\times$1~$\mu$m$^{2}$ spotsize and photon flux $I_0=10^10$ photons/sec) and the low-density device (red, 3$\times$1~$\mu$m$^{2}$ spotsize, with 3~$\mu$m horizontal width and photon flux $I_0=10^11$ photons/sec). The As fluorescence signal clearly resolves the 20~$\mu$m width of the Hall-bar and features a signal/noise ratio of $\sim7$ and $\sim2$ for the high- and low-density Hall-bars, respectively.}
    \label{fig_line_trace}
\end{figure*} 

The low-density device \#2 was also measured in an identical fashion except that the beam size was increased to $1\times3~\mu$m$^2$ yielding a photon flux of $I_0=10^{11}$~photons/sec. That way, in the low- and high-density device there were $1.7\times10^{5}$ and $1.4\times10^{6}$ As atoms within the spot size, respectively. 
Taking into account the As cross-section, the number of photons absorbed collectively by the As atoms in $200$~ms was $1.9\times10^{5}$ and $4.8\times10^{5}$ photons for the low- and high-density device, respectively.
Figure~\ref{fig_line_trace} shows the contrast obtained when measuring the As fluorescence peak intensity across the Hall-bar structures. 
The contrast in the fluorescence signal when moving the beam on and off the dopant layer has a signal/noise ratio of $\sim2$ and $\sim7$ for the low- and high-density device, respectively. Through the use of focusing elements, \textit{e.g.} Fresnel zone plates, the X-rays can be focused beyond the beam size of $\sim1~\mu$m used here. In this way, it will be possible to obtain fluorescence images of buried structures reaching a resolution of less than tens of nanometres, while still maintaining the demonstrated sensitivity to low densities \cite{Doring:13,ID16A,J.Park,Troian:2018aa}.

\section*{Weak localisation} 
The XRF images obtained from the two-dimensional As Hall-bars show that the technique is highly sensitive and directly discriminates atomic species without requiring any modelling of the sample. We measure the low-temperature electrical characteristics in magnetic fields up to $B=9$~T of the high-density Hall-bar devices \#1 before and after exposure to the X-rays.
At $T=1.8$~K electrons move diffusively, resulting in a conductivity $\sigma_0 > e^2/\hbar$ and are in the so-called weakly localised regime, as evidenced by a logarithmic temperature dependence of the zero-field conductivity \cite{disordered_electronics_ystems,HLN,altshuler_aronov,MC_silicon_inversion_layers}.
Weak localisation is a quantum interference effect that occurs for electrons in a medium with time-reversal symmetry, such as silicon, so long as the electrons' coherence length is longer than their mean free path. In that case an electron's trajectory can form a loop and interfere constructively with itself---it is weakly localised. This interference effect and particularly its behaviour in external magnetic fields depends strongly on the  disorder and dimensions of the electron channel and so is an ideal diagnostic of radiation damage. 

In the weakly localised regime, the conductance can be described by the Hikami-Larkin-Nagaoka theory \cite{HLN}, in which the conductivity change resulting from an applied magnetic field depends only on the electron mean free path $L$, the coherence length $L_\phi$, and the applied magnetic field $B$. If the conductive medium is purely two-dimensional and there are no spin-orbit or electron-electron interaction effects, only field components $B_\perp$ perpendicular to the conductive plane can couple to the electrons' orbital degree of freedom. The corresponding conductivity change is then given by
\begin{widetext}
\begin{equation}
\Delta\sigma(B_{\perp}) = \sigma_0 \left[  \psi\left(\frac{1}{2} +\frac{B_\phi}{B_{\perp}} \right) -  \psi\left(\frac{1}{2} +\frac{B_L}{B_{\perp}} \right) +\ln\left(\frac{2L_{\phi}^2}{L^2}\right) \right],
\label{eq_WL_perp}
\end{equation}
\end{widetext}
where $\psi(x)$ is the digamma function, $\sigma_0 = \frac{e^2}{2\pi^2\hbar}$, $e$ is the charge of an electron, and $\hbar$ is the reduced Planck constant. The phase breaking field is given by $B_\phi = \frac{\hbar}{4e L_{\phi}^2}$ and the elastic characteristic field $B_L =\frac{\hbar}{2eL^2}$. 

Our samples have a finite thickness, meaning that electron orbitals can have a small perpendicular component that can couple to a field $B_{||}$ parallel to the conductive plane, the effect is described by \cite{in_plane_MC}
\begin{equation}
\Delta\sigma(B_\parallel)=\sigma_0\ln(1+\gamma B_\parallel^2),
\label{eq_WL_para}
\end{equation}
where $\gamma$ is obtained by fitting the equation to the data and depends on the sample thickness $t$ and roughness. By fitting $\Delta\sigma(B_\perp)$ and $\Delta\sigma(B_\parallel)$ we can derive this thickness as \cite{sullivan} 
\begin{equation}
t=\left(\frac{1}{4\pi}\right)^{1/4}\left[\left(\frac{\hbar}{eL_{\phi}}\right)^2(\sqrt{n}L\gamma)\right]^{1/2},
\end{equation}
where $n$ is the free carrier density as measured by the Hall effect. 
Finally, for tilted magnetic fields the change in conductance can be  described by the phenomenological expression \cite{2_to_3_crossover} 
\begin{equation}
\Delta\sigma(B)^p = \Delta\sigma(B_\perp)^p+\Delta\sigma(B_\parallel)^p,
\label{WL_p}
\end{equation}
where $p$ is obtained by fitting the data and is sample and temperature-dependent.

\section*{Magneto-conductance}\label{sect-MagTrans} 
To establish whether the XRF imaging technique is non-destructive, we measure the magneto-transport at $T=1.8$~K of high-density device \#1 before and after the exposure to the X-rays.
During the X-ray imaging the sample absorbs $2'000$~photons/nm$^2$ at an energy of $11.88$~keV, corresponding to a radiation dose of $1.5\times10^{10}$~Sv ($1.5\times10^{16}$ Rad/cm$^{-2}$ or $1.7\times10^{-14}$~J/nm$^{3}$). 
Taking into account the absorption lengths of Si and the As atoms doped into silicon, as well as the As atom cross-section, we find that each As atom absorbs on average $0.3$~photons during the measurement.

\begin{figure*}[t]
  \centering
      \includegraphics[width=\linewidth,trim=0cm 2.3cm 0cm 2.8cm, clip=true]{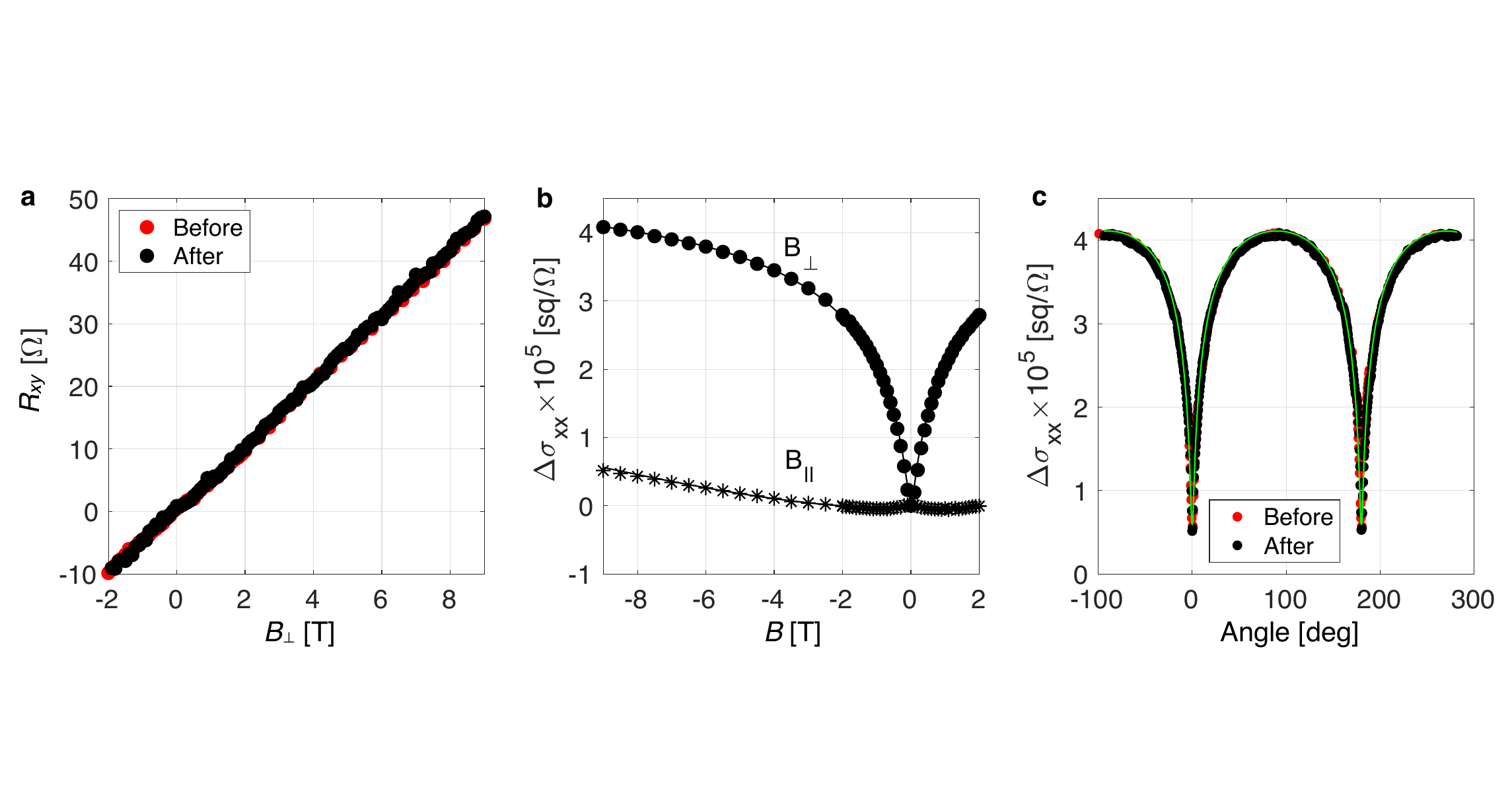}
     \caption{\textbf{Magneto-transport of the Hall-bar device before and after XRF measurements. a} Hall effect measured at $T=1.8$~K, before (red) and after (black) the XRF imaging. \textbf{b} Magneto-conductance for a magnetic field perpendicular (dots) and parallel (stars) to the conductive layer. The lines are fits to Eq. (\ref{eq_WL_perp}) and (\ref{eq_WL_para}). \textbf{c} Change of conductance in square per Ohm, as a function of the magnetic field angle with respect to the conductive layer at an external field of $B=9$~T (0 \!\!$^{\circ}\equiv$ in-plane field). The green line is a fit to Eq.(\ref{WL_p}).
	}
    \label{figMR}
\end{figure*} 

Figure \ref{figMR}a shows the Hall effect measured in a magnetic field of up to $B=9$~T, before and after the X-ray measurement shown in red and black, respectively. The transverse resistance $R_{xy}$ is linear in the field and crosses zero with no signs of quantised, non-linear, or anomalous Hall effects. Combining the Hall effect with the device's zero-field conductivity gives a mean free path $L=\sqrt{2\pi n}\mu\hbar e =4.8\pm0.1$~nm and $L=4.9\pm0.2$~nm before and after the XRF measurements, respectively, where $\mu$ is the electron mobility.
The derived Hall electron density is $n=1.31\pm0.03\times10^{14}$~cm$^{-2}$ before and $n=1.27\pm0.06\times10^{14}$~cm$^{-2}$ after the XRF measurements.
Comparing the free carrier density to the dopant density obtained from the X-ray florescence shows that the activation percentage for this device amounts to $94\pm5\%$.
Figure \ref{figMR}b shows the magneto-conductance at \mbox{$T=1.8$~K} with a field up to $B=9$~T applied perpendicular and parallel to the conductive plane. 
Fitting the data to Eq.~(\ref{eq_WL_perp}) and (\ref{eq_WL_para}) yields the electron channel's characteristic dimensions. Before the XRF, the magneto-conductance yields a coherence length of $L_\phi = 73.6\pm0.4$~nm and $\delta$ layer thickness of $t = 0.98 \pm 0.02$~nm. After the XRF measurement we obtain $L_\phi = 74.2\pm0.3$~nm and $t = 0.97 \pm 0.02$~nm.
Finally, Fig.~\ref{figMR}c shows the change in conductivity as a function of the out-of-plane angle of a 9~T magnetic field before and after the X-ray measurement. According to the Eq.~(\ref{WL_p}) the field direction-dependent data contains information of both $\Delta\sigma(B_\perp)$ and $\Delta\sigma(B_\parallel)$. Fitting the data to Eq.~(\ref{WL_p}), we obtain $p = 1.9 \pm 0.3$ and $p = 2.3 \pm 0.5$ before and after the XRF, respectively. Combining this with Eq.~(\ref{eq_WL_perp}) and (\ref{eq_WL_para}), as well as the Hall measurements implies that, within error bars, none of the device's electronic characteristics were altered by the X-ray measurement. In particular, it is important to note that the determination of the thickness by the weak-localisation measurements has a precision of 0.2~\AA, which sets a strict bound to the extent X-rays could have displaced the atoms.

\section*{Conclusions}
 We have shown that X-ray fluorescence imaging is a  technique well-suited for non-destructive investigation of dopant-based devices in silicon. This approach has the unique capability of directly identifying dopant species without relying on sample modelling, making it an attractive alternative to bb-EFM and infrared ellipsometry. X-ray scattering techniques can be used in parallel to fluorescence imaging to obtain complementary information, such as strain fields \cite{SCHULLI2018188} and overall device layout and morphology \cite{3D_imaging,high_res_3d_imaging}.
Additionally, with magneto-transport measurements, we confirm that the technique does not affect the electronic properties of the measured devices, \textit{i.e.}, it is non-destructive for Si:As, an important condition for useful device characterisation. This is in contrast to common inspection techniques such as electron microscopy and SIMS which always entail sample destruction.
Finally, with the three orders of magnitude enhancements to brilliance expected for beamlines, including focusing optics, as well as improvements both to detector solid angle and signal/noise, it is reasonable to anticipate the ability to locate single As atoms in devices to within several nm over time scales of order seconds per imaging pixel. Radiation effects will then need to be mitigated via the same strategies already exploited for X-ray ptychography today \cite{Odstrcil:2019aa,Deng:22} and diagnosed exploiting the single electron transistor characteristics of such atoms \cite{Wang:2020aa,Fuechsle:2012aa}.

\section*{Methods}\label{sect-Methods} 
\noindent\textbf{Sample preparation. }
Si(001) samples were diced to $2\times9$ mm from a 0.5~mm thick, Czochralski grown wafer, with bulk arsenic doping of density 3$\times$10$^{14}$~cm$^{-3}$, and resistivity \textgreater15~$\Omega$~cm. These samples were cleaned ultrasonically in acetone followed by isopropyl alcohol. Each sample was thermally outgassed in vacuum (base pressure \textless5$\times$10$^{-10}$~mBar) for \textgreater8~h at 600~$^{\circ}$C, and flash annealed multiple times at 1200~$^{\circ}$C, using direct current resistive sample heating. Sample temperature was monitored using an infrared pyrometer (IMPAC IGA50-LO plus), with a total estimated measurement uncertainty of $\pm$30~$^{\circ}$C.

The samples were dosed with AsH$_3$ with varying total exposures to control the dopant density. 
They were then heated at 350~$^{\circ}$C for 2~minutes to incorporate the dopants into the Si lattice \cite{Goh2005}. 
Subsequently, samples were imaged with STM, as shown in Fig.~\ref{fig_setup}, and the density of ejected Si atoms was used to estimate the density of incorporated As atoms $n_{STM}.$ 
All STM measurements were performed in an Omicron variable temperature series STM at room temperature with a base pressure of \textless5$\times$10$^{-11}$~mBar.
After incorporation, 2~nm of Si were deposited on the samples with no resistive sample heating. The samples were then resistively heated to 500~$^{\circ}$C for 15 seconds. This procedure gives a well-confined, electrically active dopant layer \cite{Keizer2015, Stock_As_count}. A further 28~nm of Si was deposited on the samples held at 250~$^{\circ}$C. Si deposition was performed at a base pressure of 2$\times$10$^{-10}$~mBar, using an all silicon, solid sublimation source (SUSI-40, MBE Komponenten GmbH) operated at a deposition rate of 0.003~nm/s. During Si deposition, the sample temperature was indirectly monitored by measuring the sample resistance, while heating using a direct current resistive sample heater.

To measure the electrical properties of the dopant layers, the samples were etched into Hall bars. This was done using optical lithography and reactive ion etching. Ohmic contacts were established by deposition of aluminium into arrays of etched holes extending through the $\delta$ layer \cite{Fuechsle2011}. 
On each sample, two Hall-bars were produced, as well as an unetched region to be used for SIMS. The samples were cleaved between the two Hall-bars. 
Each Hall-bar was mounted on a chip carrier, and electrically connected to the carrier by aluminium wire bonds.

\begin{figure*}[t]
  \centering
      \includegraphics[width=\linewidth,trim=0cm 0cm 0cm 1cm, clip=true]{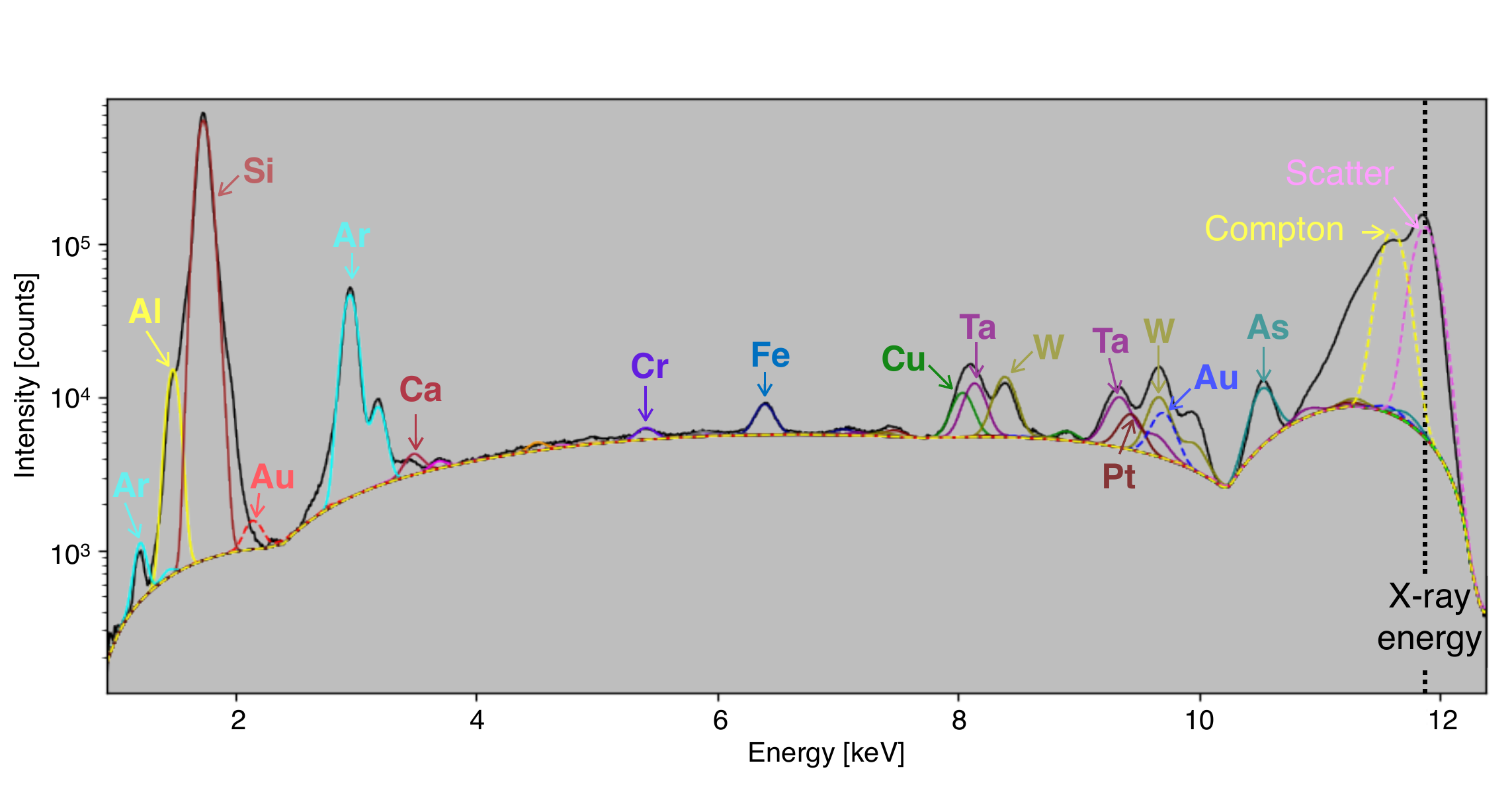}
     \caption{X-ray fluorescence spectrum taken over the highlighted area in Fig~\ref{fig_fluorescence}. The black line is the data and coloured lines are fits to one element's resonant edge. Full lines indicate a $K$-edge, while dotted lines indicate $L$ and $M$-edges as well as the scattering peaks.}
    \label{fig_spectrum}
\end{figure*}

\noindent\textbf{X-ray fluorescence spectrum. }
The fluorescence spectra obtained at each pixel of the XRF images (see Fig.~\ref{fig_fluorescence}a-c) are decomposed into a sum of elemental spectra with the help of the PyMca software \cite{PyMca}. An example of a decomposition is presented in Fig.~\ref{fig_spectrum}, where the fluorescence spectrum of each separate atomic-species is shown. Scattering peaks are also shown as dotted lines; they are the two peaks at highest energy, with the elastic scattering at the incident energy 11.88~keV and the inelastic Compton peak at slightly lower energy. 
Clearly visible in the spectrum is that each elemental spectrum contains at least one peak at a unique frequency, such that it is straightforward to identify the elements contributing to the spectrum. Only elements that cannot be excited by the incident X-ray energy cannot be detected by fluorescence. For the energy used here, 11.88~keV, the heaviest detectable element is U.
To obtain a single element image, as shown in Fig.~\ref{fig_fluorescence}a-c, it suffices to isolate in each pixel the intensity of one elemental fluorescence peak at its known energy.

\noindent\textbf{X-ray fluorescence signal/noise ratio. }
To calculate the signal/noise ratio ($SNR$) given in Fig.~\ref{fig_line_trace}, we took two 30~$\mu$m traces, one on the doped region ($T_{on}$) and one off the doped region ($T_{off}$). The SNR was then simply defined as $SNR = (\textrm{mean}(T_{On})-\textrm{mean}(T_{off}))/\textrm{std}(T_{On}))$.

\noindent\textbf{Magneto-transport setup. }
For the electrical measurements we used a standard Physical Property Measurement System (PPMS) from Quantum Design. It contains a cryostat with a superconducting magnet coil, and can control the temperature down to $T=1.8$~K and the magnetic field up to $B=9$~T. 
The samples were bonded on a standard lead-less chip carrier and inserted in a socket attached to a horizontal rotator. The rotator is motor-controlled and the rotation axis is such that the magnetic field can be set from parallel to perpendicular of the current in the Hall-bar.
The resistance is measured in a four-point geometry using a resistance bridge and with a 5~Hz square wave 100~nA current. The current is chosen to be in the linear $I-V$ response regime, and such that the Joule heating is negligible.

\noindent\textbf{Secondary-ion mass spectrometry. }
Time-of-flight SIMS measurements were conducted using an IONTOF ToF-SIMS(5) system with a 25 keV Bi$^+$ primary ion beam in high current bunch mode, and a 500 eV, 35~nA Cs$^+$ sputter beam. Depth profiles were made with a \mbox{300 $\times$ 300 $\mu$m} sputter crater, and the analytical region was the central 50 $\times$ 50 $\mu$m of the sputter region. The measured As-ion count rate was converted to a dopant density by measuring a sample of known density with the same setup.

\section*{Data availability}
The data that support the plots within this paper and other findings of this study are available from the corresponding author upon reasonable request.

\begin{acknowledgments}
We thank Jakub Vonka and Manuel Guizar Sicairos for fruitful discussions and technical assistance. We acknowledge the Paul Scherrer Institut, Villigen, Switzerland for provision of synchrotron radiation beamtime at the microXAS beamline of the Swiss Light Source. This project received funding from the European Research Council under the European Union’s Horizon 2020 research and innovation program, within the Hidden, Entangled and Resonating Order (HERO) project with Grant Agreement 810451. N.D. was partially supported by Swiss National Science Foundation Contract 175867.
We acknowledge financial support of the Engineering and Physical Sciences Research Council (EPSRC) [grant numbers EP/R034540/1, EP/W000520/1]; and Innovate UK [75574]. J.B and P.C.C. were supported by the EPSRC Centre for Doctoral Training in Advanced Characterisation of Materials (grant number EP/L015277/1), and by the Paul Scherrer Institut.
P.C.C. was partially supported by Microsoft.
\end{acknowledgments}

\section*{Author contributions}
N.D and G.A. designed the research. N.D, D.F.S., D.G. and G.A. conducted the XRF experiments and analysed the data. J.B., P.C.C., T.J.Z.S., S.F., S.R.S. and N.J.C. fabricated the devices and conducted characterisation measurements at the London Centre for Nanotechnology. N.D, M.B. and Y.S. conducted characterisation measurements at the Paul Scherrer Institut. N.D, G.M., J.B., S.G. and G.A. wrote the manuscript with input from all co-authors.

\section*{Competing interests}
The authors declare no competing interests.

\bibliographystyle{mybibstyle}
\bibliography{biblio}				
\end{document}